\documentclass[prl,twocolumn,showpacs,amsmath,amssymb]{revtex4-2}

\usepackage{hyperref}
\usepackage{graphicx}% Include figure files
\usepackage{dcolumn}% Align table columns on decimal point
\usepackage{bm}% bold math
\usepackage{subfigure}
\usepackage{amsmath}
\usepackage{amssymb}
\usepackage{cleveref}
\usepackage{color}
\usepackage{siunitx}  %for tables with decimal points very useful for clocks ...

\newcommand{\state}[3]{{}^{#1}\mathrm{#2}_{#3}}
\newcommand{\aEM}{\Delta \alpha_{\mathrm{qm}}}
\newcommand{\aE}{\Delta \alpha_{\mathrm{E1}}}
\newcommand{\ab}{\Delta \beta}
\newcommand{\Erec}{E_\mathrm{r}} 						% Photon recoil energy

\begin{document}
\title{Experimental determination of the E2-M1 polarizability of the strontium clock transition}

\author{S.~D\"orscher}
\author{J.~Klose}
\author{S.~Maratha~Palli}\altaffiliation{Present address: Active Fiber Systems GmbH, Ernst Ruska Ring 17, 07745 Jena, Germany}
\author{Ch.~Lisdat}
\email{christian.lisdat@ptb.de}
\affiliation{Physikalisch-Technische Bundesanstalt, Bundesallee 100, 38116 Braunschweig, Germany}

\date{\today}

\begin{abstract}
To operate an optical lattice clock at a fractional uncertainty below \num{e-17}, one must typically consider not only electric-dipole (E1) interaction between an atom and the lattice light field when characterizing the resulting lattice light shift of the clock transition but also higher-order multipole contributions, such as electric-quadrupole (E2) and magnetic-dipole (M1) interactions.
However, strongly incompatible values have been reported for the E2-M1 polarizability difference of the clock states $(5s5p)\,\state{3}{P}{0}$ and $(5s^2)\,\state{1}{S}{0}$ of strontium [Ushijima \textit{et al.}, Phys.\ Rev.\ Lett.\ \textbf{121} 263202 (2018); Porsev \textit{et al.}, Phys.\ Rev.\ Lett.\ \textbf{120}, 063204 (2018)].
This largely precludes operating strontium clocks with uncertainties of few \num{e-18}, as the resulting lattice light shift corrections deviate by up to \num{1e-17} from each other at typical trap depths.
We have measured the E2-M1 polarizability difference using our \textsuperscript{87}Sr lattice clock and find a value of $\aEM = -987^{+174}_{-223} \; \si{\micro\hertz}$.
This result is in very good agreement with the value reported by Ushijima \textit{et al.}
\end{abstract}

\maketitle

The interaction between the optical lattice and the trapped atom plays an important role in optical clocks with neutral atoms and has been investigated in several publications:
As the accuracy of optical lattice clocks increases, one must take into account not only the electric-dipole (E1) interaction between atom and laser field \cite{kat09} but also higher-order multipole interactions and two-photon coupling \cite{bru06, wes11, bro17, ush18, por18}. 
In electric-dipole approximation, the lattice light shift on the clock transition cancels for all lattice depths if the lattice is operated at the magic wavelength \cite{kat09}, but the higher-order contributions render this general cancellation impossible.
Lastly, the individual contributions to the lattice light shift depend intricately on the motional state of the individual atom and thus on the population distribution of the atoms in the lattice \cite{bro17, ush18, bel20}.

\begin{figure}[t]
	\includegraphics[width=\columnwidth]{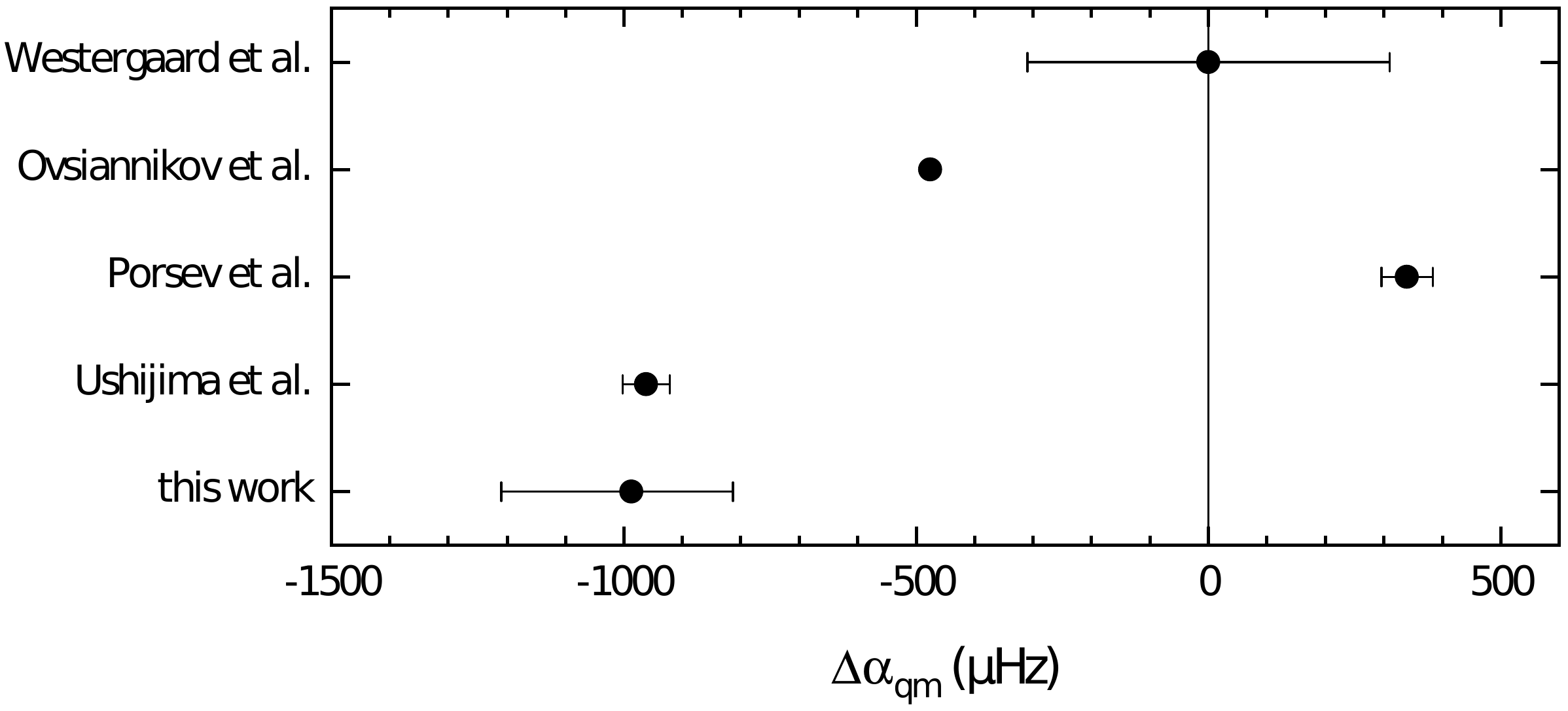}	
	\caption{Comparison of values reported for $\aEM$. (Westergaard \textit{et al.}: Ref.~\cite{wes11}, Ovsiannikov \textit{et al.}: Ref.~\cite{ovs16}, Porsev \textit{et al.}: Ref.~\cite{por18}, Ushijima \textit{et al.}: Ref.~\cite{ush18})}
	\label{fig:AlphaEM}
\end{figure}

Although the description of the light shift as a function of lattice depth can be simplified \cite{bro17}, the necessary conditions require careful testing and are not met in many cases. 
In the general case, however, several atomic parameters need to be known accurately, including the difference of the polarizabilities by electric-quadrupole (E2) and magnetic-dipole (M1) coupling, $\aEM$, at the given lattice light frequency and polarisation. 
The most accurate determinations of this atomic parameter for strontium lattice clocks have been reported by Ushijima \textit{et al.} \cite{ush18} using an experimental approach, where the different contributions to the lattice light shift are separated by their different dependences on the motional state of the atoms and on the lattice light intensity, and by Porsev \textit{et al.} \cite{por18} based on atomic structure calculations.
Worryingly, these two values are extremely incompatible with each other, as they differ by about twenty-two times their combined standard uncertainty (see Fig.~\ref{fig:AlphaEM})

\begin{figure}[tp]
	\includegraphics[width=\columnwidth]{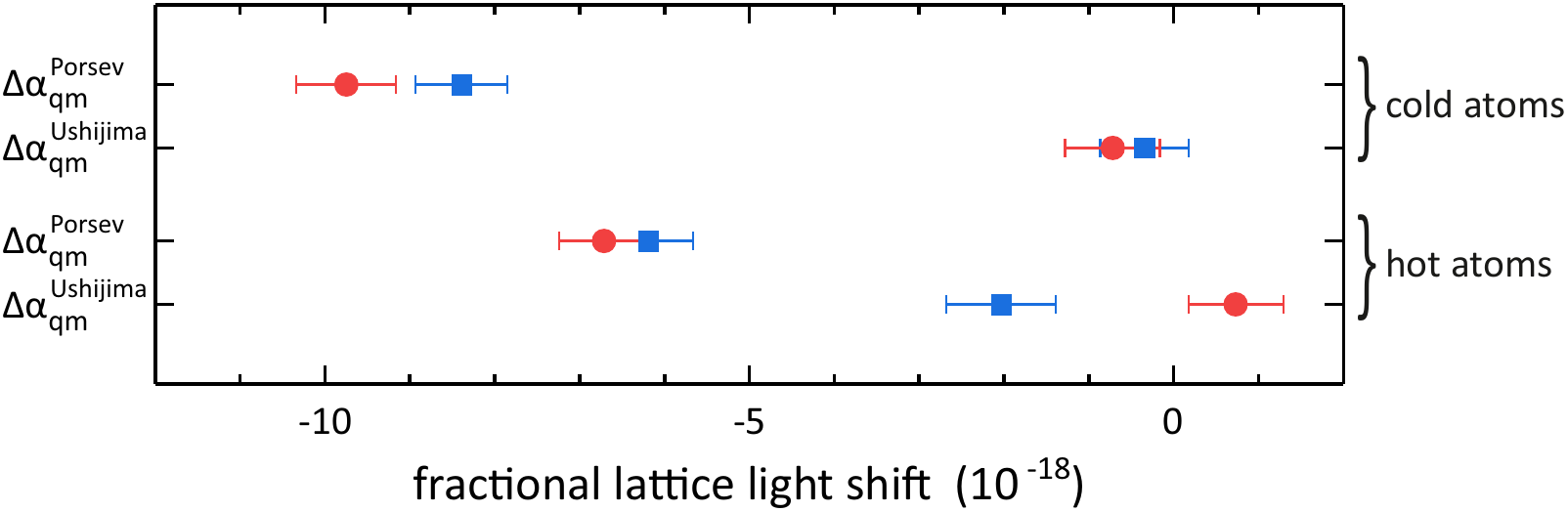}	
	\caption{
	    Lattice light shifts estimated using either the value of $\aEM$ reported by Porsev \textit{et al.} \cite{por18} or by Ushijima \textit{et al.} \cite{ush18}, for different models (dots: Ref.~\cite{bro17}; squares: Ref.~\cite{ush18}) and experimental conditions (see text) when the lattice light shift is equalized for trap depths of $77\Erec$ and $149\Erec$.
	}
	\label{fig:compMod}
\end{figure}

Given this discrepancy, it becomes difficult at best to accurately correct for the lattice light shift at an uncertainty of few \num{E-18} or less in units of the clock transition frequency (referred to as fractional units hereafter):
Using either value of $\aEM$, the E2-M1 contribution to the lattice light shift differs by about \num{1E-17} in fractional units (see Fig.~\ref{fig:compMod}) under typical conditions, including a trap depth of around $100\Erec$, where $\Erec=h^2/(2m\lambda_\textrm{m}^2)$ is the photon recoil energy at the lattice wavelength $\lambda_\textrm{m}$ for an atom of mass $m$, regardless of which light shift model \cite{bro17, ush18} is used.
Even in the motional ground state, \textit{i.e.}, in the limit of zero temperature, the difference exceeds \num{3e-18} for any reasonable \cite{lem05,aep22} lattice depth.
Hence, the discrepancy cannot be mitigated by operating at lower lattice depth or by preparing the atomic sample closer to the motional ground state, \textit{e.g.}, by cooling to sub-recoil temperatures as demonstrated recently for ytterbium \cite{zha22b}.

Here, we report on an independent experimental determination of $\aEM (\lambda_\textrm{m})$ of the clock transition in neutral strontium ($m_F = \pm 9/2$, $\Delta m_F = 0$).
Our measurement procedure follows a similar approach as the one presented in Ref.~\cite{ush18}.
We measure the differential light shift between samples with different motional state distributions at a fixed lattice depth in a new experimental apparatus that uses the same interrogation laser \cite{hae15a} as our previous system \cite{sch20d} and a vertically oriented, one-dimensional optical lattice.
The procedure used to measure differential frequency shifts is similar to those described in previous publications \cite{fal14, alm15, sch20d}, \textit{i.e.}, we run two interleaved frequency stabilization loops with different experimental conditions in the same apparatus.

\begin{figure}[tp]
	\includegraphics[width=\columnwidth]{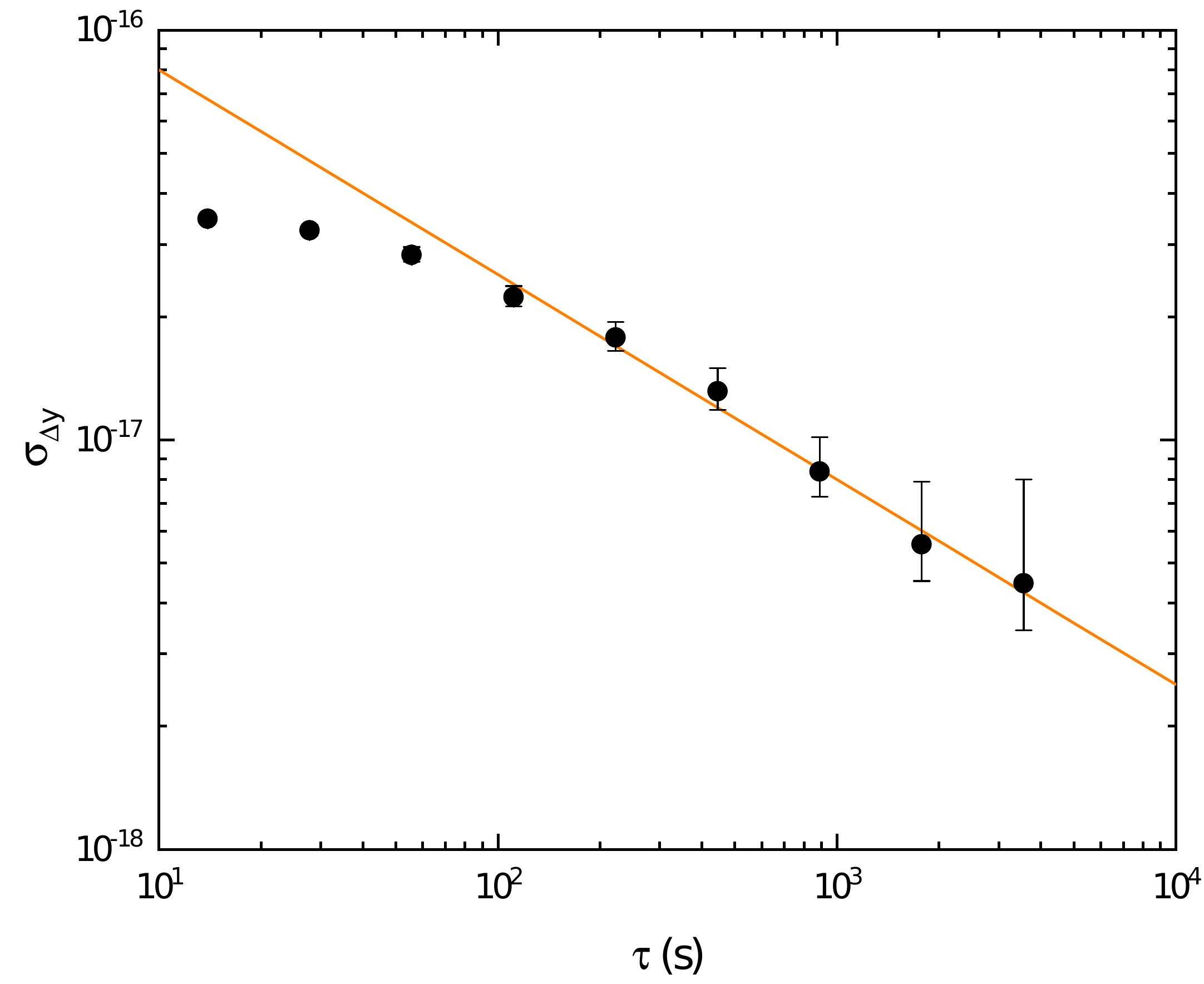}
	\caption{ Total Allan deviation of the measured differential lattice light shift between the high- and low-temperature configurations described in the main text. The line indicates an instability of $\num{2.5e-16}/\sqrt{\tau (\si{\s})}$ where $\tau$ is the measurement time.}
	\label{fig:Allan}
\end{figure}

While Ushijima \textit{et al.} compared population distributions in the axial ground state and in the first excited motional state to increase the sensitivity to $\aEM$, we induce the difference in motional state distribution by turning on or off selected cooling and filtering steps during preparation.
Following transfer of the laser-cooled atoms from the second-stage magneto-optical trap into the optical lattice at a fixed depth of about $149\Erec$, we either proceed to spectroscopy without further cooling and filtering or we transfer atoms to lower-lying axial vibrational states by sideband cooling on the $\SI{689}{\nm}$ ($\Delta F = 0$) transition and remove atoms in higher-lying vibrational states by reducing the trap depth to about $30\Erec$ for several \SI{10}{\ms} before spectroscopy at $149\Erec$ lattice depth.
The latter procedure is similar to the one described in Ref.~\cite{doe18} but uses a lower lattice depth due to the vertical orientation of the lattice beam.
Overall, this results either in a non-thermal distribution near the axial motional ground state and with strongly truncated high-energy tails in both the axial and radial degrees of freedom (`cold atoms') or in a nearly thermal distribution with substantially higher average energy in the external degrees of freedom  (`hot atoms').
We observe a differential lattice light shift of $\Delta y(\textrm{hot}-\textrm{cold}) = \num{201(24)E-19}$; the instability of this measurement is shown in Fig.~\ref{fig:Allan}.

\begin{figure}[tp]
	\includegraphics[width=\columnwidth]{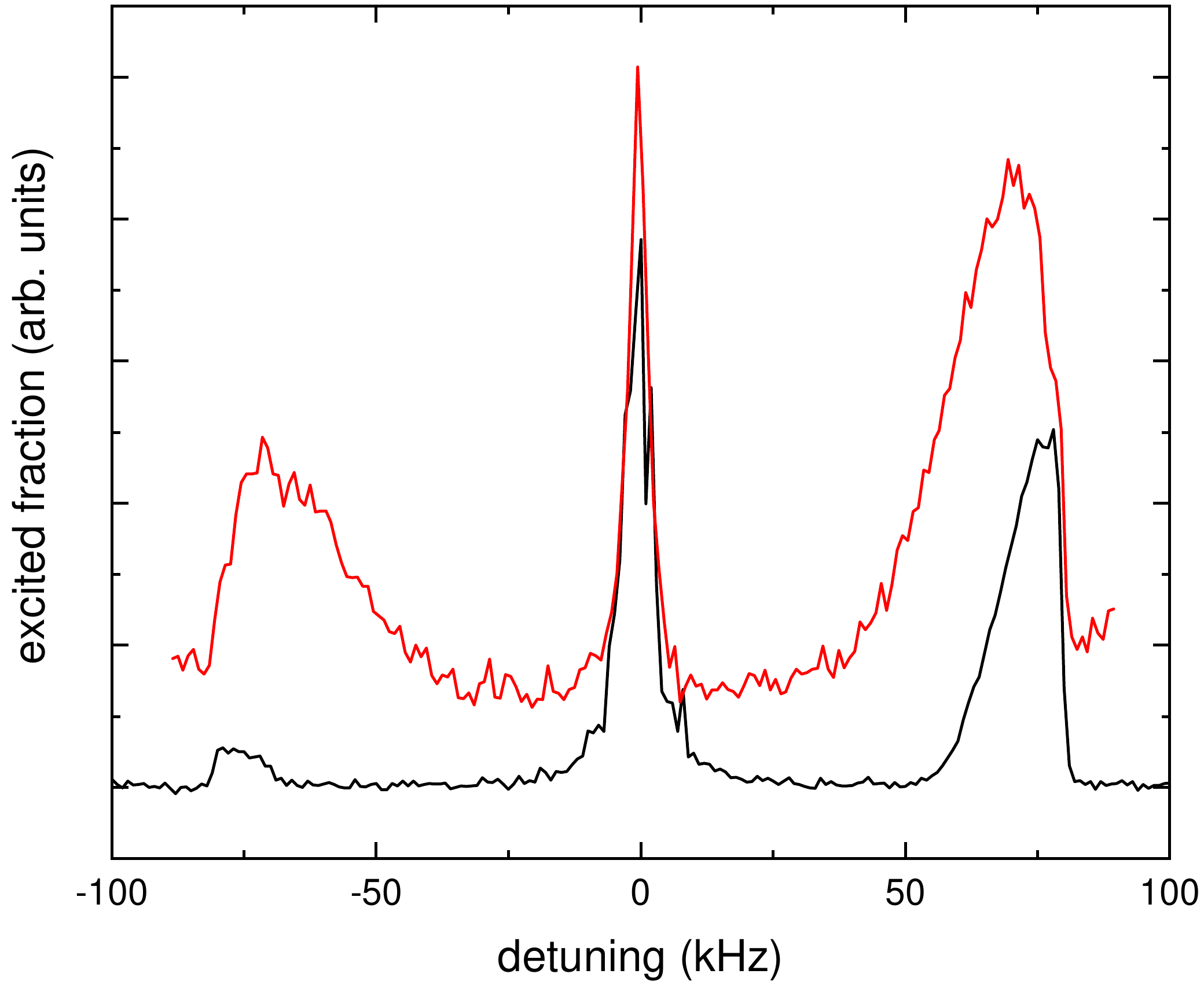}
	\caption{ Sideband spectra recorded at a lattice depth of $149\Erec$ with (lower trace) and without (upper trace) applying sideband cooling and truncating the population distribution.
	The smaller amplitude ratio of the red and blue sidebands as well as the much narrower width of the sidebands clearly show that atoms occupy lower motional states in the former case.
	We find axial (radial) temperatures of \SI{2.1}{\micro\kelvin} (\SI{5.7}{\micro\kelvin}) for `cold' atoms and \SI{6.6}{\micro\kelvin} (\SI{8.7}{\micro\kelvin}) for `hot' atoms.
	See the main text for further details.
	The curves are vertically offset for clarity.}
	\label{fig:sideband}
\end{figure}

We combine this measurement with a second measurement using the `cold' motional state distribution at two trap depths of $149\Erec$ (`hi') and $77\Erec$ (`lo') to separate the light shift's dependence on $\aEM$ from its dependence on other atomic coefficients, in particular on the differential E1 polarizability $\aE$.
We measure a differential light shift of $\Delta y(\textrm{hi} - \textrm{lo}) = \num{-173(73)E-20}$.
Finally, we determine the lattice depth and characterize the vibrational state distribution in each case using sideband spectra on the clock transition (Fig.~\ref{fig:sideband}).


We require a model of the lattice light shift to interpret the measured light shifts and extract a value for $\aEM$.
We use the model reported in Ref.~\cite{bro17} as it accounts for the dependence on both the axial and radial vibrational states and thus is well suited to describing the truncation of the population distribution, especially but not only, in the low-temperature configuration. 
The light shift as a function of the lattice depth $U$, in units of $\Erec$, is described by \cite{bro17}
\begin{eqnarray}
	\delta \nu_\mathrm{clock} &=& n_5 \aEM + \left[(n_1 + n_2) \aE - n_1\aEM \right] U^{\frac{1}{2}} \nonumber \\
		&&  - \left[ \aE + (n_3+n_4+4n_5) \ab \right] U \nonumber \\
		&& + \left[  2\ab (n_1 + n_2) \right] U^{\frac{3}{2}} - \ab U^2.
	\label{eq:LS}
\end{eqnarray}
The longitudinal ($n_z$) and radial ($n_\rho = n_x + n_y$) motional quantum numbers contribute via the factors \cite{bro17}
\begin{eqnarray*}
n_1 & = & (n_z +1/2) \mbox{,}\\
n_2 & = & [\sqrt{2}/(k w_0)](n_\rho +1) \mbox{,}\\
n_3 & = & (3/2)(n_z^2 + n_z + 1/2) \mbox{,}\\
n_4 & = & [8/(3k^2w_0^2)](n_\rho^2 + 2n_\rho + 3/2) \mbox{, and}\\
n_5 & = & 1/(\sqrt{2}k w_0)(n_z + 1/2)(n_\rho +1)\mbox{.}
\end{eqnarray*}
$k$ and $w_0$ denote the wavenumber and waist radius of the lattice, respectively.
Here, $\aEM$, $\aE$, and $\ab$ are given in frequency units for convenience.
They can be converted to their respective proper units by multiplying $\aEM$ and $\aE$ by $h \alpha_\mathrm{E1}/\Erec$ and $\ab$ by $h \left(\alpha_\mathrm{E1}/\Erec\right)^2$, where $h$ is Planck's constant and $\alpha_\mathrm{E1}$ the E1 polarizability of either clock state at the magic wavelength \cite{kat09}.
We further treat the radial degrees of freedom using the density of states given by Eq.~(3) of Ref.~\cite{bel20} for any given $n_z$, \textit{i.e.}, in the approximations of a dense energy spectrum in the radial quantum numbers $n_\rho$ and $l$ and of a harmonic trapping potential.

We model the population distribution in each case by Boltzmann distributions with effective radial and axial temperatures.
The former is derived from the shape of the respective sideband spectrum, shown in Fig.\ \ref{fig:sideband}, using the formalism of Ref.\ \cite{bla09a}, while the latter is adjusted such that the fraction of atoms in the axial vibrational ground state matches the observed ratio between the red and blue sideband amplitudes, taking into account the finite trap depth.
For `cold' atoms, the radial energy distributions for each axial vibrational state are truncated according to the reduced trap depth that is applied during preparation.
The factors $n_1$ through $n_5$ are then computed from these population distributions.

For the hyperpolarizability, the weighted average $\ab = \SI[parse-numbers = false]{458(14)}{\nano\hertz}$ of the coefficients reported in Refs.~\cite{wes11, let13, nic15, por18, ush18} is used. 
This leaves only the $\aE$ and $\aEM$ as unknown parameters in Eq.\ \ref{eq:LS}.
We can thus find the value of $\aEM$ that allows a self-consistent description of our two measurement results by Eq.~\ref{eq:LS}.
To estimate the uncertainty, we vary the respective input parameters within their uncertainties, derive the variations of $\aEM$, and add these in quadrature. 
We find
\begin{equation}
		\aEM = -987^{+174}_{-223} \; \si{\micro\hertz}\mbox{,}
	\label{eq:aEM}
\end{equation}
which is in excellent agreement with the measurement by Ushijima \textit{et al.} \cite{ush18}, but differs from the value found by Porsev \textit{et al.} \cite{por18} by more than seven times the combined standard uncertainty (Fig.~\ref{fig:AlphaEM}).

In consequence, we discard the value from Ref.~\cite{por18} and use the high-accuracy determination from Ref.~\cite{ush18}.
Then, the uncertainty contribution related to $\aEM$ in the lattice light shift determination in a strontium lattice clock is less than $\num{3E-19}$ for a trap depth of $50\Erec$ and a low-temperature configuration similar to the one described above.
In contrast, using a weighted average of the previously published data with increased uncertainty to account for the discrepancy, $\aEM = \SI[parse-numbers = false]{-370(460)}{\micro\hertz}$, would limit that uncertainty to \num{3e-18}, one order of magnitude higher. 

The data that support the findings of this work are openly available from Ref.\ [...].%\cite{doeXXX}.

We acknowledge support by the project 20FUN01 TSCAC, which has received funding from the EMPIR programme co-financed by the Participating States and from the European Union’s Horizon 2020 Research and Innovation Programme, and by the Deutsche Forschungsgemeinschaft (DFG, German Research Foundation) under Germany’s Excellence Strategy -- EXC-2123 QuantumFrontiers -- Project-ID 390837967, SFB~1464 TerraQ -- Project-ID 434617780 -- within project A04, and SFB~1227 DQ-\textit{mat} -- Project-ID 274200144 -- within project B02.
This work was partially supported by the Max Planck--RIKEN--PTB Center for Time, Constants and Fundamental Symmetries.

%%\bibliography{texbi431}

\begin{thebibliography}{19}%
\makeatletter
\providecommand \@ifxundefined [1]{%
 \@ifx{#1\undefined}
}%
\providecommand \@ifnum [1]{%
 \ifnum #1\expandafter \@firstoftwo
 \else \expandafter \@secondoftwo
 \fi
}%
\providecommand \@ifx [1]{%
 \ifx #1\expandafter \@firstoftwo
 \else \expandafter \@secondoftwo
 \fi
}%
\providecommand \natexlab [1]{#1}%
\providecommand \enquote  [1]{``#1''}%
\providecommand \bibnamefont  [1]{#1}%
\providecommand \bibfnamefont [1]{#1}%
\providecommand \citenamefont [1]{#1}%
\providecommand \href@noop [0]{\@secondoftwo}%
\providecommand \href [0]{\begingroup \@sanitize@url \@href}%
\providecommand \@href[1]{\@@startlink{#1}\@@href}%
\providecommand \@@href[1]{\endgroup#1\@@endlink}%
\providecommand \@sanitize@url [0]{\catcode `\\12\catcode `\$12\catcode
  `\&12\catcode `\#12\catcode `\^12\catcode `\_12\catcode `\%12\relax}%
\providecommand \@@startlink[1]{}%
\providecommand \@@endlink[0]{}%
\providecommand \url  [0]{\begingroup\@sanitize@url \@url }%
\providecommand \@url [1]{\endgroup\@href {#1}{\urlprefix }}%
\providecommand \urlprefix  [0]{URL }%
\providecommand \Eprint [0]{\href }%
\providecommand \doibase [0]{https://doi.org/}%
\providecommand \selectlanguage [0]{\@gobble}%
\providecommand \bibinfo  [0]{\@secondoftwo}%
\providecommand \bibfield  [0]{\@secondoftwo}%
\providecommand \translation [1]{[#1]}%
\providecommand \BibitemOpen [0]{}%
\providecommand \bibitemStop [0]{}%
\providecommand \bibitemNoStop [0]{.\EOS\space}%
\providecommand \EOS [0]{\spacefactor3000\relax}%
\providecommand \BibitemShut  [1]{\csname bibitem#1\endcsname}%
\let\auto@bib@innerbib\@empty
%</preamble>
\bibitem [{\citenamefont {Katori}\ \emph {et~al.}(2009)\citenamefont {Katori},
  \citenamefont {Hashiguchi}, \citenamefont {Il'inova},\ and\ \citenamefont
  {Ovsiannikov}}]{kat09}%
  \BibitemOpen
  \bibfield  {author} {\bibinfo {author} {\bibfnamefont {H.}~\bibnamefont
  {Katori}}, \bibinfo {author} {\bibfnamefont {K.}~\bibnamefont {Hashiguchi}},
  \bibinfo {author} {\bibfnamefont {E.~Y.}\ \bibnamefont {Il'inova}},\ and\
  \bibinfo {author} {\bibfnamefont {V.~D.}\ \bibnamefont {Ovsiannikov}},\
  }\bibfield  {title} {\bibinfo {title} {Magic wavelength to make optical
  lattice clocks insensitive to atomic motion},\ }\href
  {https://doi.org/10.1103/PhysRevLett.103.153004} {\bibfield  {journal}
  {\bibinfo  {journal} {Phys. Rev. Lett.}\ }\textbf {\bibinfo {volume} {103}},\
  \bibinfo {pages} {153004} (\bibinfo {year} {2009})}\BibitemShut {NoStop}%
\bibitem [{\citenamefont {Brusch}\ \emph {et~al.}(2006)\citenamefont {Brusch},
  \citenamefont {Le~Targat}, \citenamefont {Baillard}, \citenamefont
  {Fouch\'e},\ and\ \citenamefont {Lemonde}}]{bru06}%
  \BibitemOpen
  \bibfield  {author} {\bibinfo {author} {\bibfnamefont {A.}~\bibnamefont
  {Brusch}}, \bibinfo {author} {\bibfnamefont {R.}~\bibnamefont {Le~Targat}},
  \bibinfo {author} {\bibfnamefont {X.}~\bibnamefont {Baillard}}, \bibinfo
  {author} {\bibfnamefont {M.}~\bibnamefont {Fouch\'e}},\ and\ \bibinfo
  {author} {\bibfnamefont {P.}~\bibnamefont {Lemonde}},\ }\bibfield  {title}
  {\bibinfo {title} {Hyperpolarizability effects in a {Sr} optical lattice
  clock},\ }\href {http://link.aps.org/abstract/PRL/v96/e103003} {\bibfield
  {journal} {\bibinfo  {journal} {Phys. Rev. Lett.}\ }\textbf {\bibinfo
  {volume} {96}},\ \bibinfo {pages} {103003} (\bibinfo {year}
  {2006})}\BibitemShut {NoStop}%
\bibitem [{\citenamefont {Westergaard}\ \emph {et~al.}(2011)\citenamefont
  {Westergaard}, \citenamefont {Lodewyck}, \citenamefont {Lorini},
  \citenamefont {Lecallier}, \citenamefont {Burt}, \citenamefont {Zawada},
  \citenamefont {Millo},\ and\ \citenamefont {Lemonde}}]{wes11}%
  \BibitemOpen
  \bibfield  {author} {\bibinfo {author} {\bibfnamefont {P.~G.}\ \bibnamefont
  {Westergaard}}, \bibinfo {author} {\bibfnamefont {J.}~\bibnamefont
  {Lodewyck}}, \bibinfo {author} {\bibfnamefont {L.}~\bibnamefont {Lorini}},
  \bibinfo {author} {\bibfnamefont {A.}~\bibnamefont {Lecallier}}, \bibinfo
  {author} {\bibfnamefont {E.~A.}\ \bibnamefont {Burt}}, \bibinfo {author}
  {\bibfnamefont {M.}~\bibnamefont {Zawada}}, \bibinfo {author} {\bibfnamefont
  {J.}~\bibnamefont {Millo}},\ and\ \bibinfo {author} {\bibfnamefont
  {P.}~\bibnamefont {Lemonde}},\ }\bibfield  {title} {\bibinfo {title}
  {Lattice-induced frequency shifts in {Sr} optical lattice clocks at the
  $10^{-17}$ level},\ }\href {https://doi.org/10.1103/PhysRevLett.106.210801}
  {\bibfield  {journal} {\bibinfo  {journal} {Phys. Rev. Lett.}\ }\textbf
  {\bibinfo {volume} {106}},\ \bibinfo {pages} {210801} (\bibinfo {year}
  {2011})}\BibitemShut {NoStop}%
\bibitem [{\citenamefont {Brown}\ \emph {et~al.}(2017)\citenamefont {Brown},
  \citenamefont {Phillips}, \citenamefont {Beloy}, \citenamefont {McGrew},
  \citenamefont {Schioppo}, \citenamefont {Fasano}, \citenamefont {Milani},
  \citenamefont {Zhang}, \citenamefont {Hinkley}, \citenamefont {Leopardi},
  \citenamefont {Yoon}, \citenamefont {Nicolodi}, \citenamefont {Fortier},\
  and\ \citenamefont {Ludlow}}]{bro17}%
  \BibitemOpen
  \bibfield  {author} {\bibinfo {author} {\bibfnamefont {R.~C.}\ \bibnamefont
  {Brown}}, \bibinfo {author} {\bibfnamefont {N.~B.}\ \bibnamefont {Phillips}},
  \bibinfo {author} {\bibfnamefont {K.}~\bibnamefont {Beloy}}, \bibinfo
  {author} {\bibfnamefont {W.~F.}\ \bibnamefont {McGrew}}, \bibinfo {author}
  {\bibfnamefont {M.}~\bibnamefont {Schioppo}}, \bibinfo {author}
  {\bibfnamefont {R.~J.}\ \bibnamefont {Fasano}}, \bibinfo {author}
  {\bibfnamefont {G.}~\bibnamefont {Milani}}, \bibinfo {author} {\bibfnamefont
  {X.}~\bibnamefont {Zhang}}, \bibinfo {author} {\bibfnamefont
  {N.}~\bibnamefont {Hinkley}}, \bibinfo {author} {\bibfnamefont
  {H.}~\bibnamefont {Leopardi}}, \bibinfo {author} {\bibfnamefont {T.~H.}\
  \bibnamefont {Yoon}}, \bibinfo {author} {\bibfnamefont {D.}~\bibnamefont
  {Nicolodi}}, \bibinfo {author} {\bibfnamefont {T.~M.}\ \bibnamefont
  {Fortier}},\ and\ \bibinfo {author} {\bibfnamefont {A.~D.}\ \bibnamefont
  {Ludlow}},\ }\bibfield  {title} {\bibinfo {title} {Hyperpolarizability and
  operational magic wavelength in an optical lattice clock},\ }\href
  {https://doi.org/10.1103/PhysRevLett.119.253001} {\bibfield  {journal}
  {\bibinfo  {journal} {Phys. Rev. Lett.}\ }\textbf {\bibinfo {volume} {119}},\
  \bibinfo {pages} {253001} (\bibinfo {year} {2017})}\BibitemShut {NoStop}%
\bibitem [{\citenamefont {Ushijima}\ \emph {et~al.}(2018)\citenamefont
  {Ushijima}, \citenamefont {Takamoto},\ and\ \citenamefont {Katori}}]{ush18}%
  \BibitemOpen
  \bibfield  {author} {\bibinfo {author} {\bibfnamefont {I.}~\bibnamefont
  {Ushijima}}, \bibinfo {author} {\bibfnamefont {M.}~\bibnamefont {Takamoto}},\
  and\ \bibinfo {author} {\bibfnamefont {H.}~\bibnamefont {Katori}},\
  }\bibfield  {title} {\bibinfo {title} {Operational magic intensity for {Sr}
  optical lattice clocks},\ }\href
  {https://doi.org/10.1103/PhysRevLett.121.263202} {\bibfield  {journal}
  {\bibinfo  {journal} {Phys. Rev. Lett.}\ }\textbf {\bibinfo {volume} {121}},\
  \bibinfo {pages} {263202} (\bibinfo {year} {2018})}\BibitemShut {NoStop}%
\bibitem [{\citenamefont {Porsev}\ \emph {et~al.}(2018)\citenamefont {Porsev},
  \citenamefont {Safronova}, \citenamefont {Safronova},\ and\ \citenamefont
  {Kozlov}}]{por18}%
  \BibitemOpen
  \bibfield  {author} {\bibinfo {author} {\bibfnamefont {S.~G.}~\bibnamefont
  {Porsev}}, \bibinfo {author} {\bibfnamefont {M.~S.}~\bibnamefont {Safronova}},
  \bibinfo {author} {\bibfnamefont {U~.I.}~\bibnamefont {Safronova}},\ and\
  \bibinfo {author} {\bibfnamefont {M.~G.}~\bibnamefont {Kozlov}},\ }\bibfield
  {title} {\bibinfo {title} {Multipolar polarizabilities and
  hyperpolarizabilities in the {Sr} optical lattice clock},\ }\href
  {https://doi.org/10.1103/PhysRevLett.120.063204} {\bibfield  {journal}
  {\bibinfo  {journal} {Phys. Rev. Lett.}\ }\textbf {\bibinfo {volume} {120}},\
  \bibinfo {pages} {063204} (\bibinfo {year} {2018})}\BibitemShut {NoStop}%
\bibitem [{\citenamefont {Beloy}\ \emph {et~al.}(2020)\citenamefont {Beloy},
  \citenamefont {McGrew}, \citenamefont {Zhang}, \citenamefont {Nicolodi},
  \citenamefont {Fasano}, \citenamefont {Hassan}, \citenamefont {Brown},\ and\
  \citenamefont {Ludlow}}]{bel20}%
  \BibitemOpen
  \bibfield  {author} {\bibinfo {author} {\bibfnamefont {K.}~\bibnamefont
  {Beloy}}, \bibinfo {author} {\bibfnamefont {W.~F.}\ \bibnamefont {McGrew}},
  \bibinfo {author} {\bibfnamefont {X.}~\bibnamefont {Zhang}}, \bibinfo
  {author} {\bibfnamefont {D.}~\bibnamefont {Nicolodi}}, \bibinfo {author}
  {\bibfnamefont {R.~J.}\ \bibnamefont {Fasano}}, \bibinfo {author}
  {\bibfnamefont {Y.~S.}\ \bibnamefont {Hassan}}, \bibinfo {author}
  {\bibfnamefont {R.~C.}\ \bibnamefont {Brown}},\ and\ \bibinfo {author}
  {\bibfnamefont {A.~D.}\ \bibnamefont {Ludlow}},\ }\bibfield  {title}
  {\bibinfo {title} {Modeling motional energy spectra and lattice light shifts
  in optical lattice clocks},\ }\href
  {https://doi.org/10.1103/PhysRevA.101.053416} {\bibfield  {journal} {\bibinfo
   {journal} {Phys. Rev. A}\ }\textbf {\bibinfo {volume} {101}},\ \bibinfo
  {pages} {053416} (\bibinfo {year} {2020})}\BibitemShut {NoStop}%
\bibitem [{\citenamefont {Lemonde}\ and\ \citenamefont {Wolf}(2005)}]{lem05}%
  \BibitemOpen
  \bibfield  {author} {\bibinfo {author} {\bibfnamefont {P.}~\bibnamefont
  {Lemonde}}\ and\ \bibinfo {author} {\bibfnamefont {P.}~\bibnamefont {Wolf}},\
  }\bibfield  {title} {\bibinfo {title} {Optical lattice clock with atoms
  confined in a shallow trap},\ }\href
  {https://doi.org/10.1103/PhysRevA.72.033409} {\bibfield  {journal} {\bibinfo
  {journal} {Phys. Rev. A}\ }\textbf {\bibinfo {volume} {72}},\ \bibinfo
  {pages} {033409} (\bibinfo {year} {2005})}\BibitemShut {NoStop}%
\bibitem [{\citenamefont {Aeppli}\ \emph {et~al.}(2022)\citenamefont {Aeppli},
  \citenamefont {Chu}, \citenamefont {Bothwell}, \citenamefont {Kennedy},
  \citenamefont {Kedar}, \citenamefont {He}, \citenamefont {Rey},\ and\
  \citenamefont {Ye}}]{aep22}%
  \BibitemOpen
  \bibfield  {author} {\bibinfo {author} {\bibfnamefont {A.}~\bibnamefont
  {Aeppli}}, \bibinfo {author} {\bibfnamefont {A.}~\bibnamefont {Chu}},
  \bibinfo {author} {\bibfnamefont {T.}~\bibnamefont {Bothwell}}, \bibinfo
  {author} {\bibfnamefont {C.~J.}\ \bibnamefont {Kennedy}}, \bibinfo {author}
  {\bibfnamefont {D.}~\bibnamefont {Kedar}}, \bibinfo {author} {\bibfnamefont
  {P.}~\bibnamefont {He}}, \bibinfo {author} {\bibfnamefont {A.~M.}\
  \bibnamefont {Rey}},\ and\ \bibinfo {author} {\bibfnamefont {J.}~\bibnamefont
  {Ye}},\ }\bibfield  {title} {\bibinfo {title} {Hamiltonian engineering of
  spin-orbit--coupled fermions in a {Wannier-Stark} optical lattice clock},\
  }\href {https://doi.org/10.1126/sciadv.adc9242} {\bibfield  {journal}
  {\bibinfo  {journal} {Science Advances}\ }\textbf {\bibinfo {volume} {8}},\
  \bibinfo {pages} {eadc9242} (\bibinfo {year} {2022})},\ \Eprint
  {https://arxiv.org/abs/https://www.science.org/doi/pdf/10.1126/sciadv.adc9242}
  {https://www.science.org/doi/pdf/10.1126/sciadv.adc9242} \BibitemShut
  {NoStop}%
\bibitem [{\citenamefont {Zhang}\ \emph {et~al.}(2022)\citenamefont {Zhang},
  \citenamefont {Beloy}, \citenamefont {Hassan}, \citenamefont {McGrew},
  \citenamefont {Chen}, \citenamefont {Siegel}, \citenamefont {Grogan},\ and\
  \citenamefont {Ludlow}}]{zha22b}%
  \BibitemOpen
  \bibfield  {author} {\bibinfo {author} {\bibfnamefont {X.}~\bibnamefont
  {Zhang}}, \bibinfo {author} {\bibfnamefont {K.}~\bibnamefont {Beloy}},
  \bibinfo {author} {\bibfnamefont {Y.~S.}\ \bibnamefont {Hassan}}, \bibinfo
  {author} {\bibfnamefont {W.~F.}\ \bibnamefont {McGrew}}, \bibinfo {author}
  {\bibfnamefont {C.-C.}\ \bibnamefont {Chen}}, \bibinfo {author}
  {\bibfnamefont {J.~L.}\ \bibnamefont {Siegel}}, \bibinfo {author}
  {\bibfnamefont {T.}~\bibnamefont {Grogan}},\ and\ \bibinfo {author}
  {\bibfnamefont {A.~D.}\ \bibnamefont {Ludlow}},\ }\bibfield  {title}
  {\bibinfo {title} {Subrecoil clock-transition laser cooling enabling shallow
  optical lattice clocks},\ }\href
  {https://doi.org/10.1103/PhysRevLett.129.113202} {\bibfield  {journal}
  {\bibinfo  {journal} {Phys. Rev. Lett.}\ }\textbf {\bibinfo {volume} {129}},\
  \bibinfo {pages} {113202} (\bibinfo {year} {2022})}\BibitemShut {NoStop}%
\bibitem [{\citenamefont {H{\"a}fner}\ \emph {et~al.}(2015)\citenamefont
  {H{\"a}fner}, \citenamefont {Falke}, \citenamefont {Grebing}, \citenamefont
  {Vogt}, \citenamefont {Legero}, \citenamefont {Merimaa}, \citenamefont
  {Lisdat},\ and\ \citenamefont {Sterr}}]{hae15a}%
  \BibitemOpen
  \bibfield  {author} {\bibinfo {author} {\bibfnamefont {S.}~\bibnamefont
  {H{\"a}fner}}, \bibinfo {author} {\bibfnamefont {S.}~\bibnamefont {Falke}},
  \bibinfo {author} {\bibfnamefont {C.}~\bibnamefont {Grebing}}, \bibinfo
  {author} {\bibfnamefont {S.}~\bibnamefont {Vogt}}, \bibinfo {author}
  {\bibfnamefont {T.}~\bibnamefont {Legero}}, \bibinfo {author} {\bibfnamefont
  {M.}~\bibnamefont {Merimaa}}, \bibinfo {author} {\bibfnamefont
  {C.}~\bibnamefont {Lisdat}},\ and\ \bibinfo {author} {\bibfnamefont
  {U.}~\bibnamefont {Sterr}},\ }\bibfield  {title} {\bibinfo {title} {$8 \times
  10^{-17}$ fractional laser frequency instability with a long room-temperature
  cavity},\ }\href {https://doi.org/10.1364/OL.40.002112} {\bibfield  {journal}
  {\bibinfo  {journal} {Opt. Lett.}\ }\textbf {\bibinfo {volume} {40}},\
  \bibinfo {pages} {2112} (\bibinfo {year} {2015})}\BibitemShut {NoStop}%
\bibitem [{\citenamefont {Schwarz}\ \emph {et~al.}(2020)\citenamefont
  {Schwarz}, \citenamefont {D\"{o}rscher}, \citenamefont {Al-Masoudi},
  \citenamefont {Benkler}, \citenamefont {Legero}, \citenamefont {Sterr},
  \citenamefont {Weyers}, \citenamefont {Rahm}, \citenamefont {Lipphardt},\
  and\ \citenamefont {Lisdat}}]{sch20d}%
  \BibitemOpen
  \bibfield  {author} {\bibinfo {author} {\bibfnamefont {R.}~\bibnamefont
  {Schwarz}}, \bibinfo {author} {\bibfnamefont {S.}~\bibnamefont
  {D\"{o}rscher}}, \bibinfo {author} {\bibfnamefont {A.}~\bibnamefont
  {Al-Masoudi}}, \bibinfo {author} {\bibfnamefont {E.}~\bibnamefont {Benkler}},
  \bibinfo {author} {\bibfnamefont {T.}~\bibnamefont {Legero}}, \bibinfo
  {author} {\bibfnamefont {U.}~\bibnamefont {Sterr}}, \bibinfo {author}
  {\bibfnamefont {S.}~\bibnamefont {Weyers}}, \bibinfo {author} {\bibfnamefont
  {J.}~\bibnamefont {Rahm}}, \bibinfo {author} {\bibfnamefont {B.}~\bibnamefont
  {Lipphardt}},\ and\ \bibinfo {author} {\bibfnamefont {C.}~\bibnamefont
  {Lisdat}},\ }\bibfield  {title} {\bibinfo {title} {Long term measurement of
  the {$^{87}$Sr} clock frequency at the limit of primary {Cs} clocks},\ }\href
  {https://doi.org/10.1103/PhysRevResearch.2.033242} {\bibfield  {journal}
  {\bibinfo  {journal} {Phys. Rev. Research}\ }\textbf {\bibinfo {volume}
  {2}},\ \bibinfo {pages} {033242} (\bibinfo {year} {2020})}\BibitemShut
  {NoStop}%
\bibitem [{\citenamefont {Falke}\ \emph {et~al.}(2014)\citenamefont {Falke},
  \citenamefont {Lemke}, \citenamefont {Grebing}, \citenamefont {Lipphardt},
  \citenamefont {Weyers}, \citenamefont {Gerginov}, \citenamefont {Huntemann},
  \citenamefont {Hagemann}, \citenamefont {Al-Masoudi}, \citenamefont
  {H{\"a}fner}, \citenamefont {Vogt}, \citenamefont {Sterr},\ and\
  \citenamefont {Lisdat}}]{fal14}%
  \BibitemOpen
  \bibfield  {author} {\bibinfo {author} {\bibfnamefont {S.}~\bibnamefont
  {Falke}}, \bibinfo {author} {\bibfnamefont {N.}~\bibnamefont {Lemke}},
  \bibinfo {author} {\bibfnamefont {C.}~\bibnamefont {Grebing}}, \bibinfo
  {author} {\bibfnamefont {B.}~\bibnamefont {Lipphardt}}, \bibinfo {author}
  {\bibfnamefont {S.}~\bibnamefont {Weyers}}, \bibinfo {author} {\bibfnamefont
  {V.}~\bibnamefont {Gerginov}}, \bibinfo {author} {\bibfnamefont
  {N.}~\bibnamefont {Huntemann}}, \bibinfo {author} {\bibfnamefont
  {C.}~\bibnamefont {Hagemann}}, \bibinfo {author} {\bibfnamefont
  {A.}~\bibnamefont {Al-Masoudi}}, \bibinfo {author} {\bibfnamefont
  {S.}~\bibnamefont {H{\"a}fner}}, \bibinfo {author} {\bibfnamefont
  {S.}~\bibnamefont {Vogt}}, \bibinfo {author} {\bibfnamefont {U.}~\bibnamefont
  {Sterr}},\ and\ \bibinfo {author} {\bibfnamefont {C.}~\bibnamefont
  {Lisdat}},\ }\bibfield  {title} {\bibinfo {title} {A strontium lattice clock
  with $3 \times 10^{-17}$ inaccuracy and its frequency},\ }\href
  {https://doi.org/10.1088/1367-2630/16/7/073023} {\bibfield  {journal}
  {\bibinfo  {journal} {New J. Phys.}\ }\textbf {\bibinfo {volume} {16}},\
  \bibinfo {pages} {073023} (\bibinfo {year} {2014})}\BibitemShut {NoStop}%
\bibitem [{\citenamefont {Al-Masoudi}\ \emph {et~al.}(2015)\citenamefont
  {Al-Masoudi}, \citenamefont {D\"orscher}, \citenamefont {H\"afner},
  \citenamefont {Sterr},\ and\ \citenamefont {Lisdat}}]{alm15}%
  \BibitemOpen
  \bibfield  {author} {\bibinfo {author} {\bibfnamefont {A.}~\bibnamefont
  {Al-Masoudi}}, \bibinfo {author} {\bibfnamefont {S.}~\bibnamefont
  {D\"orscher}}, \bibinfo {author} {\bibfnamefont {S.}~\bibnamefont
  {H\"afner}}, \bibinfo {author} {\bibfnamefont {U.}~\bibnamefont {Sterr}},\
  and\ \bibinfo {author} {\bibfnamefont {C.}~\bibnamefont {Lisdat}},\
  }\bibfield  {title} {\bibinfo {title} {Noise and instability of an optical
  lattice clock},\ }\href {https://doi.org/10.1103/PhysRevA.92.063814}
  {\bibfield  {journal} {\bibinfo  {journal} {Phys. Rev. A}\ }\textbf {\bibinfo
  {volume} {92}},\ \bibinfo {pages} {063814} (\bibinfo {year}
  {2015})}\BibitemShut {NoStop}%
\bibitem [{\citenamefont {D{\"o}rscher}\ \emph {et~al.}(2018)\citenamefont
  {D{\"o}rscher}, \citenamefont {Schwarz}, \citenamefont {Al-Masoudi},
  \citenamefont {Falke}, \citenamefont {Sterr},\ and\ \citenamefont
  {Lisdat}}]{doe18}%
  \BibitemOpen
  \bibfield  {author} {\bibinfo {author} {\bibfnamefont {S.}~\bibnamefont
  {D{\"o}rscher}}, \bibinfo {author} {\bibfnamefont {R.}~\bibnamefont
  {Schwarz}}, \bibinfo {author} {\bibfnamefont {A.}~\bibnamefont {Al-Masoudi}},
  \bibinfo {author} {\bibfnamefont {S.}~\bibnamefont {Falke}}, \bibinfo
  {author} {\bibfnamefont {U.}~\bibnamefont {Sterr}},\ and\ \bibinfo {author}
  {\bibfnamefont {C.}~\bibnamefont {Lisdat}},\ }\bibfield  {title} {\bibinfo
  {title} {Lattice-induced photon scattering in an optical lattice clock},\
  }\href {https://doi.org/10.1103/PhysRevA.97.063419} {\bibfield  {journal}
  {\bibinfo  {journal} {Phys. Rev. A}\ }\textbf {\bibinfo {volume} {97}},\
  \bibinfo {pages} {063419} (\bibinfo {year} {2018})}\BibitemShut {NoStop}%
\bibitem [{\citenamefont {Blatt}\ \emph {et~al.}(2009)\citenamefont {Blatt},
  \citenamefont {Thomsen}, \citenamefont {Campbell}, \citenamefont {Ludlow},
  \citenamefont {Swallows}, \citenamefont {Martin}, \citenamefont {Boyd},\ and\
  \citenamefont {Ye}}]{bla09a}%
  \BibitemOpen
  \bibfield  {author} {\bibinfo {author} {\bibfnamefont {S.}~\bibnamefont
  {Blatt}}, \bibinfo {author} {\bibfnamefont {J.~W.}\ \bibnamefont {Thomsen}},
  \bibinfo {author} {\bibfnamefont {G.~K.}\ \bibnamefont {Campbell}}, \bibinfo
  {author} {\bibfnamefont {A.~D.}\ \bibnamefont {Ludlow}}, \bibinfo {author}
  {\bibfnamefont {M.~D.}\ \bibnamefont {Swallows}}, \bibinfo {author}
  {\bibfnamefont {M.~J.}\ \bibnamefont {Martin}}, \bibinfo {author}
  {\bibfnamefont {M.~M.}\ \bibnamefont {Boyd}},\ and\ \bibinfo {author}
  {\bibfnamefont {J.}~\bibnamefont {Ye}},\ }\bibfield  {title} {\bibinfo
  {title} {Rabi spectroscopy and excitation inhomogeneity in a one-dimensional
  optical lattice clock},\ }\href {https://doi.org/10.1103/PhysRevA.80.052703}
  {\bibfield  {journal} {\bibinfo  {journal} {Phys. Rev. A}\ }\textbf {\bibinfo
  {volume} {80}},\ \bibinfo {pages} {052703} (\bibinfo {year}
  {2009})}\BibitemShut {NoStop}%
\bibitem [{\citenamefont {Le~Targat}\ \emph {et~al.}(2013)\citenamefont
  {Le~Targat}, \citenamefont {Lorini}, \citenamefont {Le~Coq}, \citenamefont
  {Zawada}, \citenamefont {Gu\'ena}, \citenamefont {Abgrall}, \citenamefont
  {Gurov}, \citenamefont {Rosenbusch}, \citenamefont {Rovera}, \citenamefont
  {Nag\'orny}, \citenamefont {Gartman}, \citenamefont {Westergaard},
  \citenamefont {Tobar}, \citenamefont {Lours}, \citenamefont {Santarelli},
  \citenamefont {Clairon}, \citenamefont {Bize}, \citenamefont {Laurent},
  \citenamefont {Lemonde},\ and\ \citenamefont {Lodewyck}}]{let13}%
  \BibitemOpen
  \bibfield  {author} {\bibinfo {author} {\bibfnamefont {R.}~\bibnamefont
  {Le~Targat}}, \bibinfo {author} {\bibfnamefont {L.}~\bibnamefont {Lorini}},
  \bibinfo {author} {\bibfnamefont {Y.}~\bibnamefont {Le~Coq}}, \bibinfo
  {author} {\bibfnamefont {M.}~\bibnamefont {Zawada}}, \bibinfo {author}
  {\bibfnamefont {J.}~\bibnamefont {Gu\'ena}}, \bibinfo {author} {\bibfnamefont
  {M.}~\bibnamefont {Abgrall}}, \bibinfo {author} {\bibfnamefont
  {M.}~\bibnamefont {Gurov}}, \bibinfo {author} {\bibfnamefont
  {P.}~\bibnamefont {Rosenbusch}}, \bibinfo {author} {\bibfnamefont {D.~G.}\
  \bibnamefont {Rovera}}, \bibinfo {author} {\bibfnamefont {B.}~\bibnamefont
  {Nag\'orny}}, \bibinfo {author} {\bibfnamefont {R.}~\bibnamefont {Gartman}},
  \bibinfo {author} {\bibfnamefont {P.~G.}\ \bibnamefont {Westergaard}},
  \bibinfo {author} {\bibfnamefont {M.~E.}\ \bibnamefont {Tobar}}, \bibinfo
  {author} {\bibfnamefont {M.}~\bibnamefont {Lours}}, \bibinfo {author}
  {\bibfnamefont {G.}~\bibnamefont {Santarelli}}, \bibinfo {author}
  {\bibfnamefont {A.}~\bibnamefont {Clairon}}, \bibinfo {author} {\bibfnamefont
  {S.}~\bibnamefont {Bize}}, \bibinfo {author} {\bibfnamefont {P.}~\bibnamefont
  {Laurent}}, \bibinfo {author} {\bibfnamefont {P.}~\bibnamefont {Lemonde}},\
  and\ \bibinfo {author} {\bibfnamefont {J.}~\bibnamefont {Lodewyck}},\
  }\bibfield  {title} {\bibinfo {title} {Experimental realization of an optical
  second with strontium lattice clocks},\ }\href
  {https://doi.org/10.1038/ncomms3109} {\bibfield  {journal} {\bibinfo
  {journal} {Nature Commun.}\ }\textbf {\bibinfo {volume} {4}},\ \bibinfo
  {pages} {2109} (\bibinfo {year} {2013})}\BibitemShut {NoStop}%
\bibitem [{\citenamefont {Nicholson}\ \emph {et~al.}(2015)\citenamefont
  {Nicholson}, \citenamefont {Campbell}, \citenamefont {Hutson}, \citenamefont
  {Marti}, \citenamefont {Bloom}, \citenamefont {McNally}, \citenamefont
  {Zhang}, \citenamefont {Barrett}, \citenamefont {Safronova}, \citenamefont
  {Strouse}, \citenamefont {Tew},\ and\ \citenamefont {Ye}}]{nic15}%
  \BibitemOpen
  \bibfield  {author} {\bibinfo {author} {\bibfnamefont {T.~L.}\ \bibnamefont
  {Nicholson}}, \bibinfo {author} {\bibfnamefont {S.~L.}\ \bibnamefont
  {Campbell}}, \bibinfo {author} {\bibfnamefont {R.~B.}\ \bibnamefont
  {Hutson}}, \bibinfo {author} {\bibfnamefont {G.~E.}\ \bibnamefont {Marti}},
  \bibinfo {author} {\bibfnamefont {B.~J.}\ \bibnamefont {Bloom}}, \bibinfo
  {author} {\bibfnamefont {R.~L.}\ \bibnamefont {McNally}}, \bibinfo {author}
  {\bibfnamefont {W.}~\bibnamefont {Zhang}}, \bibinfo {author} {\bibfnamefont
  {M.~D.}\ \bibnamefont {Barrett}}, \bibinfo {author} {\bibfnamefont {M.~S.}\
  \bibnamefont {Safronova}}, \bibinfo {author} {\bibfnamefont {G.~F.}\
  \bibnamefont {Strouse}}, \bibinfo {author} {\bibfnamefont {W.~L.}\
  \bibnamefont {Tew}},\ and\ \bibinfo {author} {\bibfnamefont {J.}~\bibnamefont
  {Ye}},\ }\bibfield  {title} {\bibinfo {title} {Systematic evaluation of an
  atomic clock at $2 \times 10^{-18}$ total uncertainty},\ }\href
  {https://doi.org/10.1038/ncomms7896} {\bibfield  {journal} {\bibinfo
  {journal} {Nature Commun.}\ }\textbf {\bibinfo {volume} {6}},\ \bibinfo
  {pages} {6896} (\bibinfo {year} {2015})}\BibitemShut {NoStop}%
\bibitem [{\citenamefont {Ovsiannikov}\ \emph {et~al.}(2016)\citenamefont
  {Ovsiannikov}, \citenamefont {Marmo}, \citenamefont {Palchikov},\ and\
  \citenamefont {Katori}}]{ovs16}%
  \BibitemOpen
  \bibfield  {author} {\bibinfo {author} {\bibfnamefont {V.~D.}\ \bibnamefont
  {Ovsiannikov}}, \bibinfo {author} {\bibfnamefont {S.~I.}\ \bibnamefont
  {Marmo}}, \bibinfo {author} {\bibfnamefont {V.~G.}\ \bibnamefont
  {Palchikov}},\ and\ \bibinfo {author} {\bibfnamefont {H.}~\bibnamefont
  {Katori}},\ }\bibfield  {title} {\bibinfo {title} {Higher-order effects on
  the precision of clocks of neutral atoms in optical lattices},\ }\href
  {https://doi.org/10.1103/PhysRevA.93.043420} {\bibfield  {journal} {\bibinfo
  {journal} {Phys. Rev. A}\ }\textbf {\bibinfo {volume} {93}},\ \bibinfo
  {pages} {043420} (\bibinfo {year} {2016})}\BibitemShut {NoStop}%
\end{thebibliography}

\end{document}